\documentclass[twocolumn,superscriptaddress,aps,prb]{revtex4}
\usepackage{graphicx}
\usepackage{color}

\begin{document}
\title{Topological phase transition and two dimensional topological insulators in Ge-based thin films}
\author{Bahadur Singh}
\affiliation{Department of Physics, Indian Institute of Technology Kanpur, Kanpur 208016, India}

\author{Hsin Lin}
\affiliation{Department of Physics, Northeastern University, Boston, Massachusetts 02115, USA} 
\affiliation{Graphene Research Centre and Department of Physics, National University of Singapore, Singapore 117542} 

\author{R. Prasad}
\affiliation{Department of Physics, Indian Institute of Technology Kanpur, Kanpur 208016, India}

\author{A. Bansil}
\affiliation{Department of Physics, Northeastern University, Boston, Massachusetts 02115, USA}

\begin{abstract}
We discuss possible topological phase transitions in Ge-based thin films of Ge(Bi$_x$Sb$_{1-x}$)$_2$Te$_4$ as a function of layer thickness and Bi concentration $x$ using the first principles
density functional theory framework. The bulk material is a topological insulator at $x$ = 1.0 with a single Dirac cone surface state at the surface Brillouin zone center, whereas it is a trivial insulator 
at $x$ = 0. Through a systematic examination of the band topologies we predict that thin films of Ge(Bi$_x$Sb$_{1-x}$)$_2$Te$_4$ with $x$ = 0.6, 0.8 and 1.0 are candidates for two-dimensional 
(2D) topological insulators, which would undergo a 2D topological phase transition as a function of $x$. A topological phase diagram for Ge(Bi$_x$Sb$_{1-x}$)$_2$Te$_4$ thin films is 
presented to help guide their experimental exploration.
\end{abstract}

 
\maketitle

\section{Introduction}
Topological insulators (TIs) are novel materials in which even though the bulk system is insulating, the surface can support spin-polarized gapless states with Dirac-cone-like linear
energy dispersion.\cite{qi,hasan,moore} The topological surface states are unique in being robust against scattering from non-magnetic impurities, and display spin-momentum locking, which
results in helical spin textures. \cite{qshe,bervenig} TIs not only offer exciting possibilities for applications in spintronics, energy and information technologies, but also provide 
platforms for exploring in a solid state setting questions which have traditionally been considered to lie in the realm of high energy physics, such as the Weyl semimetal phases and the 
Higgs mechanism.\cite{wan,burkov,halasz,murakami,singh,higgs,mass} 

Two dimensional (2D) topological insulators, also referred to as the quantum spin Hall (QSH) insulators, were predicted theoretically, before being realized experimentally in HgTe/CdTe 
quantum wells.\cite{bervenig,konig} The three-dimensional (3D) TIs were identified later in bismuth-based thermoelectrics, Bi$_{1-x}$Sb$_x$, Bi$_2$Se$_3$, Bi$_2$Te$_3$, and 
Sb$_2$Te$_3$,\cite{zhang,hsieh,xia} although transport properties of these binary TIs are dominated by intrinsic vacancies and disorder in the bulk material. By now a variety of 3D 
TIs have been proposed theoretically and verified experimentally in a number of cases. \cite{lin,tetra,neupane,eremeev_pbNC,heuslar,throughput} In sharp contrast, to date, the only
experimental realizations of the QSH state are HgTe/CdTe and InAs/GaSb/AlSb quantum well systems.\cite{InAsGaSbtheory,InAsGaSbexp} No standalone thin film or a thin film supported on a 
suitable substrate has been realized as a QSH state, although various theoretical proposals have been made suggesting that 2D TIs could be achieved through the reduced dimensionality in 
thin films of 3D TIs.\cite{lin,silkin,liubi2te3,yzhangnat584,APL-Sb} The need for finding new QSH insulator materials is for these reasons obvious.

A topological phase transition (TPT) from a trivial to a non-trivial topological phase in 2D is an interesting unexplored issue, although in 3D a TPT has been demonstrated in TlBi(Se,S)$_2$ 
solid solutions.\cite{mass} Despite the theoretical prediction for the existence of nontrivial 2D TIs in this family of materials\cite{lin}, no experimental realization has been reported, 
which may be due to stronger bonding in the Tl-compounds compared to the weaker van der Waals type bonding between quintuple layers in the Bi$_2$Se$_3$ family. Interestingly, rhombohedral 
Sb$_2$Se$_3$ has been predicted to be a trivial insulator, implying that a TPT could be realized in (Bi$_{1-x}$Sb$_{x}$)$_2$Se$_3$ solid solutions. However, the real Sb$_2$Se$_3$ material 
exhibits an orthorhombic structure, and a structural phase transition intervenes before the TPT point is reached, as the Sb concentration increases.

These considerations suggest that a strategy for realizing a thin-film material exhibiting a 2D TPT is to begin with an existing 3D topological material in which the layers are weakly bonded
and trivial and non-trivial topological phases can be achieved without encountering a structural instability. Here we recall that GeBi$_2$Te$_4$ (GBT124) was theoretically predicted 
\cite{eremmev_jept,arxiv} and experimentally verified\cite{okamoto,neupane} as a 3D TI, but GeSb$_2$Te$_4$ (GST124) is predicted \cite{arxiv} to be a trivial insulator with a large bulk 
band-gap. GBT124 and GST124 thus support different topological characters, but possess a similar lattice structure.
Notably, insulating samples of Bi$_2$Te$_3$ and Bi$_2$Se$_3$ have proven difficult to realize experimentally, but this problem may prove more tractable in the Ge(Bi$_x$Sb$_{1-x}$)$_2$Te$_4$ 
system. In particular, the Dirac point in GBT124, for example, is well isolated from the bulk bands,\cite{neupane,okamoto} while it lies very close to the valence band maximum or it is
buried in the bulk bands in Bi$_2$Te$_3$/Bi$_2$Se$_3$. Along this line, experimental studies of GBT124 show a larger surface state spin-polarization ($\sim$70\%) \cite{arxiv,okamoto}
compared to Bi$_2$Te$_3$ and Bi$_2$Se$_3$ ($\sim$50\%-60\%).\cite{prl_plolar} Also, being ternary compounds, there is greater flexibility in substitutions, for example, the Ge site in 
GBT124 can be replaced by Sn or Pb to tune the lattice constant and electronic structure. GBT124 and GST124 thus are good candidate parent compounds for investigating a TPT. Also, if we
consider related 2D slabs thinner than the surface state decay length\cite{zhou_l,wada}, these slabs may yield a material supporting a 2D TI as well as a 2D 
TPT.\cite{silkin,liubi2te3,yzhangnat584} Moreover, since GBT124 is known to be $n$-type and topologically nontrivial, adding Sb could reduce electron carriers, leading to a more insulating
compound. Accordingly, this paper examines the evolution of topological characteristics of Ge(Bi$_x$Sb$_{1-x}$)$_2$Te$_4$ by systematically varying the concentration $x$ of Bi atoms for 
various layer thicknesses. We find N-layer (NL) films of Ge(Bi$_x$Sb$_{1-x}$)$_2$Te$_4$ to be 2D TIs as follows: 28L and 35L films at $x$ = 0.6; 21L, 28L, and 35L films at $x$ = 0.8; and 14L,
21L and 28L films at $x$ = 1.0. The material undergoes a 3D phase transition to topological regime near $x$ = 0.6. We have also constructed a topological phase diagram for 2D 
Ge(Bi$_x$Sb$_{1-x}$)$_2$Te$_4$ thin films with varying thickness and Bi-concentrations. 

The present article is organized as follows. Section II gives computational details. In Sec. III, we delineate the bulk crystal and electronic structures of GBT124 and GST124. 
The TPT in Ge(Bi$_x$Sb$_{1-x}$)$_2$Te$_4$ with varying $x$ and the existence of 2D TIs in thin films of Ge(Bi$_x$Sb$_{1-x}$)$_2$Te$_4$ with various concentrations are also discussed. 
Finally, Sec. IV summarizes our conclusions.

\section{Computational Details }
Our electronic structure calculations were carried out within the density functional theory (DFT)\cite{kohan} framework with the projector-augmented-wave (PAW) basis\cite{vasp,paw}. The 
generalized-gradient approximation (GGA)\cite{pbe} was used to include exchange-correlation effects, and the spin-orbit coupling (SOC) effects were included as implemented in VASP 
(Vienna Ab Initio Simulation Package).\cite{vasp} The bulk calculations used a plane wave cut-off energy of 350 eV and a $\Gamma$-centered 8$\times$8$\times$8 k-mesh with conjugate gradient 
algorithm\cite{cga}. Since Sb and Bi atoms possess a similar outermost electronic configuration, evolution of electronic structure could be tracked by varying either Bi or Sb concentration, 
and we have chosen to do so by varying the concentration of Bi atoms for the sake of definiteness. Specifically, a bulk hexagonal supercell with 35-atomic layers (five septuple layers) was 
prepared with values of $x$ varying between 0.0 and 1.0. The corresponding bulk parameters, including the structure were optimized, until all components of Hellman-Feynman forces on each 
ion were less than 0.001 eV/\r{A}. The relaxed structures of Ge(Bi$_x$Sb$_{1-x}$)$_2$Te$_4$ examined for various Bi concentrations were also found to be hexagonal. 
The surface electronic-structure calculations are based on a slab geometry with a plane wave cut-off energy of 350 eV, a $\Gamma$-centered 8$\times$8$\times$1 k-mesh, and relaxed bulk 
parameters with a vacuum greater than 12 \r{A}.

\section{Results and Discussions}
\subsection{Bulk crystal and band structure}

The GBT124 and GST124 belong to the rhombohedral crystal structure\cite{neupane,okamoto,arxiv}, composed of seven-layer (7L) or septuple blocks, with layers in the  sequence 
Te-Bi(Sb)-Te-Ge-Te-Bi(Sb)-Te. As an example, the arrangement of layers in a unit cell in GBT124 is shown in Fig. 1(a). In a septuple block, the Ge and Bi atoms are sandwiched between 
the Te-atoms, and the Ge atom can be chosen as the inversion center. The bonding within the septuple blocks is strong, being mainly of ionic-covalent type, whereas across the septuple
blocks bonding is of van der Waals type. \cite{okamoto,neupane} With this in mind, we take the surface termination to occur between two septuple blocks and regard one septuple block as 
a 2D thin film with the smallest thickness considered. 

\begin{figure}[ht!] 
\includegraphics[width=0.45\textwidth]{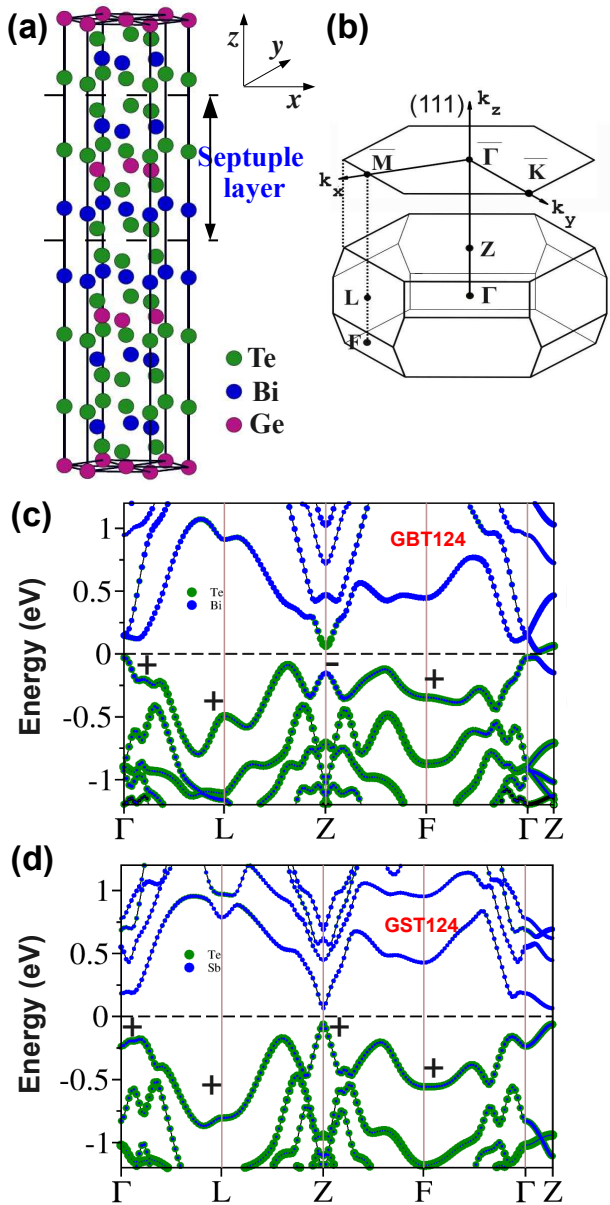} 
\caption {(Color online) (a) Bulk crystal structure of GeBi$_2$Te$_4$ (GBT124) with inversion-symmetry. A seven atomic layer block is shown. (b) The corresponding bulk Brillouin zone (BZ) with
four time reversal invariant points $\Gamma$, $F$, $Z$ and $L$, and the 2D Brillouin zone of the (111) surface with two time reversal invariant points $\overline{\Gamma}$ and $\overline{M}$.
(c-d) The bulk electronic structure of GeBi$_2$Te$_4$(GBT124) and GeSb$_2$Te$_4$(GST124), respectively with Bi/Sb (blue dots) and Te (green dots) atomic weights for different bands.
Signs of $\delta{_i}=\pm 1$ at TRIM are also shown.}
\end{figure}

The bulk band structures of GBT124 and GST124 are shown in Figs. 1(c) and 1(d), respectively. GBT124 is an indirect band-gap semiconductor in which the conduction band minimum (CBM) and 
valence band maxima (VBM) lie along the $\Gamma-$Z direction. The bulk valence and conduction bands at the $Z$ point are composed of Bi $p$ states and Te $p$ states, respectively, with an 
inverted band order. In contrast, GST124 is a direct band-gap material with CBM and VBM at $Z$ point and a normal ordering of bands. Since the structure possesses inversion symmetry in both 
cases, it is possible to calculate the $Z_2$ invariants $\nu_0$;($\nu_1 \nu_2 \nu_3$) (where $\nu_0$ is a strong and $\nu_{k=1,2,3}$ is a weak topological invariant)\cite{kane} from the bulk 
band structure. The $Z_2$ invariants are determined from the parity $\xi_m(\Gamma_i)$ of the occupied bulk bands at the time 
reversal invariant momentum (TRIM) points $\Gamma_{i=(n_1n_2n_3)}$ = ($n_1b_1+n_2b_2+n_3b_3$)/2, where b$_{1}$, b$_{2}$, and b$_{3}$ are the reciprocal lattice vectors and n$_k$ = 0 or 1.\cite{kane}
The $Z_2$ invariants can then be calculated using
\begin{equation}
 (-1)^{\nu_0}=\prod_{i=1}^8\delta_i
\end{equation}
and
\begin{equation}
 (-1)^{\nu_k}=\prod_{n_k=1;n_{j\ne k}=0,1}\delta_{i=(n_1,n_2,n_3)}
\end{equation}
where
\begin{equation}
 \delta_i=\prod_{m=1}^N \xi_{2m}(\Gamma_i)
\end{equation}
Here, N is the number of occupied bulk bands and $\xi_{2m}(\Gamma_i)=\pm{1}$ is the parity of the 2m$^{th}$ occupied energy band at the point $\Gamma_i$.
There are eight TRIM points in the rhombohedral Brillouin zone, but only four of these points [$\Gamma$, Z, F and L; see Fig. 1(b)], are inequivalent. 
The product of the parity eigenvalues ($\delta{_i}=\pm1$) of the occupied bands at the TRIM points are shown in Figs. 1(c)-1(d). In the case of GBT124, interestingly, the band inversion occurs
at the Z-point, which leads to $Z_2$ invariants being equal to 1;(111), and is different from Bi$_2$Se$_3$ TI family\cite{zhang}, where the band inversion occurs at the $\Gamma$-point, 
yielding $Z_2$ invariants to be 1;(000). On other hand, GST124 lacks band inversion at any of the TRIM points, indicating that the system is a normal insulator with all $Z_2$ invariants zero.

\subsection{ Surface band structures and 3D topological phase transition}
For investigating the TPT in Ge(Bi$_x$Sb$_{1-x}$)$_2$Te$_4$ compounds, we used a supercell geometry with a relaxed hexagonal supercell having 35 atomic layers for $x$ = 0.0, 0.2, 0.4, 0.5, 
0.6, 0.8 and 1.0.\cite{footnoteC, disorder, VCA} As examples, the bulk hexagonal supercells for $x$=1.0 and $x$ = 0.8 are shown in Fig.~2. For $x$ = 1.0, the Ge atom in the supercell remains 
an inversion center, whereas for $x$ = 0.8 this is not the case and the inversion symmetry is broken.

\begin{figure}[ht!] 
\includegraphics[width=0.45\textwidth]{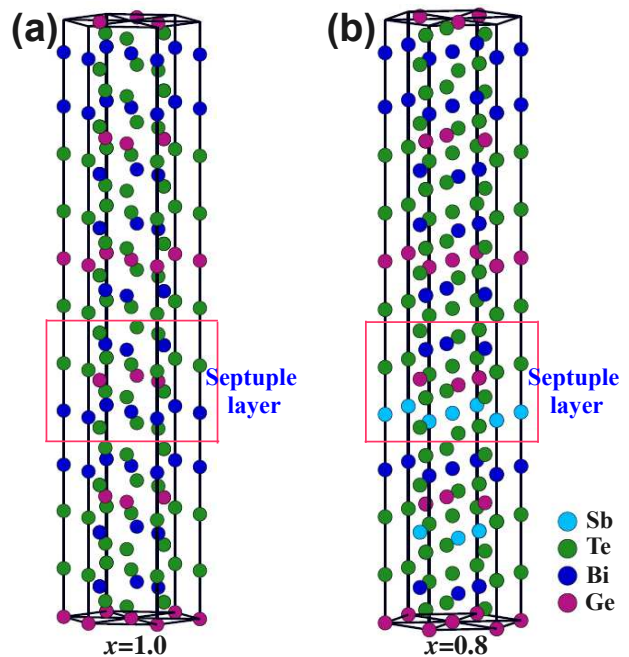}
\caption{(Color online) Schematic diagram of the bulk hexagonal supercell with 35 atomic layers for (a) $x$ = 1.0 (with inversion symmetry) and  (b) $x$ = 0.8 (without inversion symmetry). 
Red boxes identify the septuple layer arrangement.}
\end{figure}

The bulk band structures were computed using the fully relaxed structures. 
Figure 3(a) shows the
variation of bulk band gap of Ge(Bi$_x$Sb$_{1-x}$)$_2$Te$_4$ as a function of $x$. The gap is seen to start decreasing from 0.13 eV at $x$ = 0.0, with the valence band and conduction band 
mainly composed of Te p-states and Bi/Sb p-states, respectively [Fig.~3(b)], to a minimum value around 0.5$<x<$0.6. As we further increase $x$, the gap opens up again and attains a value
of $\approx$0.07 eV at $x$=1.0 with valence and conduction bands swapping their orbital characters at the $Z$-point. This closing and reopening of the bulk gap with
an inverted band order indicates that there is a TPT between $x$ = 0.5 to $x$ = 0.6.\cite{footnoteA, footnoteB}

\begin{figure*}[ht!] 
\includegraphics[width=1.0\textwidth]{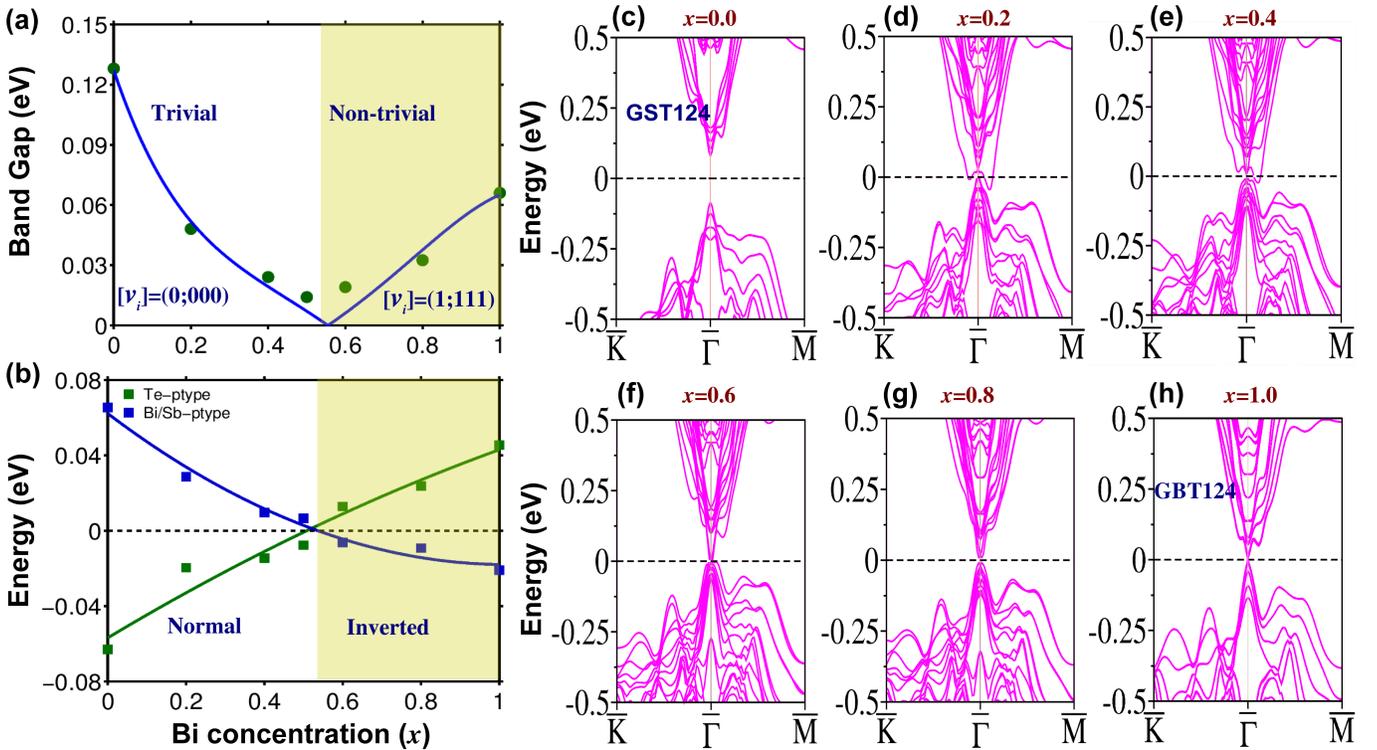}
\caption{(Color online) (a) Variation of band gap of bulk hexagonal supercell of Ge(Bi$_x$Sb$_{1-x}$)$_2$Te$_4$ with $x$. The related topological invariants are shown. Blue line is a guide to the eye. 
(b) Evolution of Te p-type (green points) and Bi/Sb p-type (blue points) valence band maximum and conduction band minimum at the $Z$-point with $x$. The valence and conduction bands are seen
to be inverted between 0.5 $<x<$ 0.6. Green and blue lines are guides to the eye. Panels (c)$-$(h) show the surface electronic structures for various $x$ values. There is a phase transition 
from normal insulator (NI) to topological insulator (TI) near $x$ = 0.6, although a small gap is seen at $x$ = 0.6 and 0.8, making these compositions suitable candidates for realizing 2D TIs.
The dashed zero lines mark the Fermi energy.}
\end{figure*}

Figures. 3(c) -3(h) show electronic structures of 35-atomic-layer slabs for various Bi concentrations. Figure 3(c) shows that the $x$ = 0.0 compound GST124 is a normal insulator with a large 
band gap without a gapless surface state inside the bulk energy gap region, which is consistent with the trivial insulator found in bulk calculations. On the other hand, a clear Dirac cone
surface state at the $\bar\Gamma$ point is seen in Fig. 3(h) for the nontrivial phase GBT124 at $x$=1. The system becomes metallic at $x$ = 0.2 and 0.4 due to a surface conduction
band crossing the Fermi level. Between the $\bar\Gamma$ point and the $\bar{M}$ point, this surface band crosses the Fermi level twice, an even number, consistent with the trivial phase
found in bulk calculations. The metallic character decreases with increasing $x$ and at $x$ = 0.6 and 0.8, it becomes insulating with a very small gap ($\approx$ 10 and 18 meV at $x$ = 0.6 
and $x$ = 0.8, respectively). This gap is due to quantum confinement effects, i.e., the interaction between the two Dirac cones residing on the top and bottom surfaces. Without such interaction,
each side of the surface has gapless Dirac-cone surface states centered at the $\bar\Gamma$ point, which arises from the nontrivial bulk band topology.

\subsection{2D topological insulators}
We now turn to discuss the evolution of electronic structure with slab thickness in terms of multiples of septuple layer blocks. Reference ~\onlinecite{wada} has previously shown that when the 
thickness of a slab is smaller than the surface state decay length, states on the two surfaces of the slab become coupled via quantum tunneling, leading to a small thickness-dependent gap in
the electronic structure.\cite{singh,park,yzhangnat584} This coupling between the two surfaces for thin slabs is responsible for opening a gap at the Dirac point and is the key for 
realizing the insulating phase. Since thin films of Ge(Bi$_x$Sb$_{1-x}$)$_2$Te$_4$ with $x$ = 0.0 (GST124) as well as $x$ = 1.0 (GBT124)[Fig. 2(a)] are symmetric under inversion, we used parity 
analysis \cite{kane} to determine their topological character. On the other hand, thin films for $x$ = 0.2, 0.4, 0.5, 0.6 and 0.8 are asymmetric under inversion [Fig. 2(b)], 
and therefore parity-analysis cannot be used. Instead, we varied the SOC strength and monitor the band gap to assess the topological character. 
We have further verified this non-trivial character via edge state computations.
 
\begin{figure}[ht!]
\includegraphics[width=0.45\textwidth]{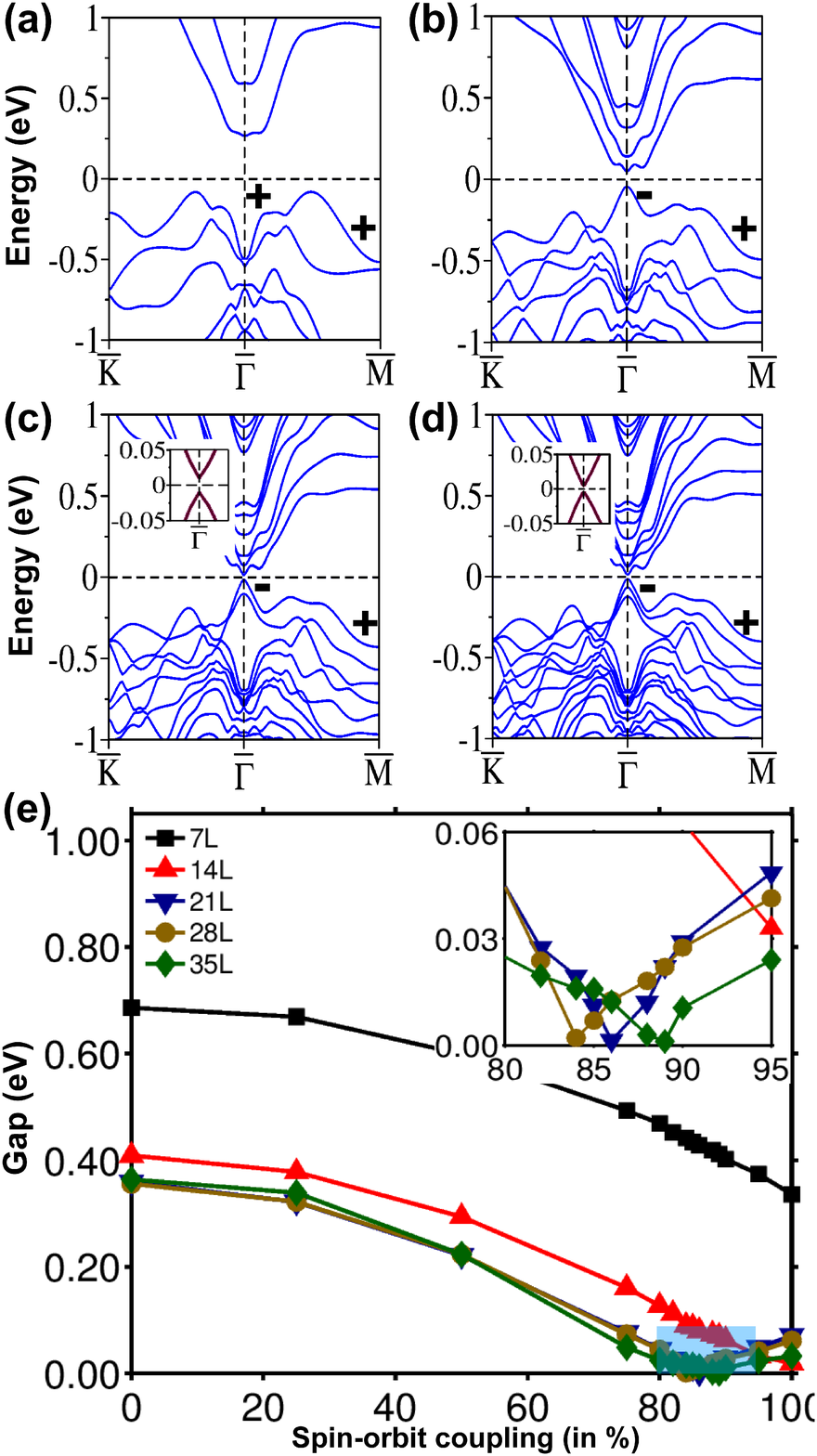} 
\caption{(Color online) Electronic structure of Ge(Bi$_x$Sb$_{1-x}$)$_2$Te$_4$ at $x$ = 1.0 for (a) 7L, (b) 14L, (c) 21L, and (d) 28L thick films. Signs of $\delta{_i}=\pm 1$ at the TRIM points of the 2D Brillouin zone
are also shown. Panel (e) shows the variation of band gap for various film thicknesses in Ge(Bi$_x$Sb$_{1-x}$)$_2$Te$_4$ at $x$ = 0.8 as a function of the size of spin-orbit coupling. The
inset zooms in on the shaded region. The energy gap decreases to zero and reopens as SOC increases from zero to 100\% for 21L, 28L, and 35L slabs.}
\end{figure}

Figures 4(a)-(d) show electronic structures of GBT124 films ($x$ = 1.0) for various thicknesses. It is evident that the slab with 7L displays an indirect band gap, whereas thicker slabs with 14L,
21L, and 28L have a direct band gap. Since all these slabs are insulating, we examined the possibility of these slabs being 2D TIs. Note that slabs with $x$ = 1.0 are symmetric under inversion,
so that the Bloch wave functions have well defined parity at the TRIM points, which can be used to compute the 2D topological invariants, which are also shown in Figs. 4(a)-(d). The 2D $Z_2$ 
invariant assumes the nontrivial value equal to 1 for 14L, 21L, and 28L films, but has the trivial value of zero for the 7L film. 

As already pointed out, thin films with $x$ = 0.2, 0.4, 0.5, 0.6, and 0.8 are asymmetric under inversion. In order to check their topological character, we varied the SOC strength from zero 
to 100\%. Since the topological phase for all insulators without SOC is trivial, we can monitor the gap size to determine if a TPT takes place as the strength of the SOC is increased. Results for
$x$ = 0.8 are shown in Fig. 4(e). For 7L and 14L films the band gap decreases with increasing values of the SOC strength, being 0.336 eV for 7L and 0.019 eV for 14L at 100\% SOC without closing
at any value of SOC strength. Therefore, band structures with and without SOC are adiabatically connected, implying that they are both topologically trivial. On the other hand, for 21L, 28L
and 35L films, the gap closes at 86\%, 84\%, and 89\% SOC strength, respectively, and re-opens to a value of 0.071 eV for 21L, 0.062 eV for 28L, and 0.032 eV for the 35L film at 100\% SOC 
strength. We thus conclude that the slabs of Ge(Bi$_x$Sb$_{1-x}$)$_2$Te$_4$ at $x$ = 0.8, with 21L, 28L and 35L thickness are topologically non-trivial.

\begin{figure}[h!]
\includegraphics[width=0.45\textwidth]{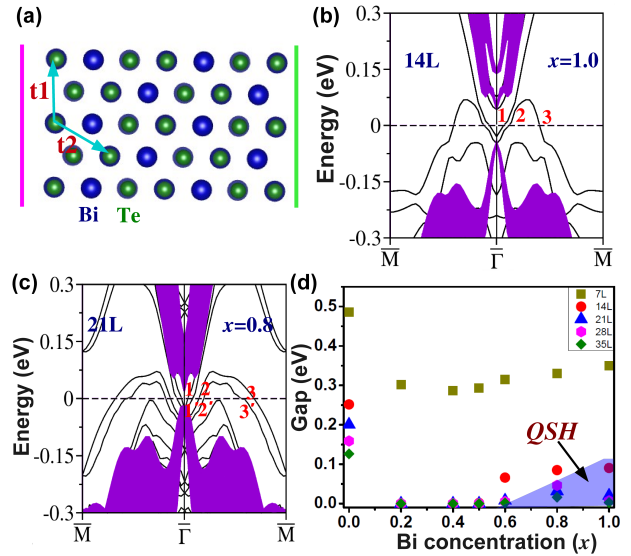} 
\caption{(Color online) (a)Top view of 2D thin films of GBT124 with in-plane lattice vectors \textbf{t1} and \textbf{t2}. The edge states have been computed along the edges shown by pink and green lines. 
The edge states for: (b) 14L GBT124, and (c) for 21L Ge(Bi$_x$Sb$_{1-x}$)$_2$Te$_4$ at $x$ = 0.8. The continuum of bands is shown in violet color. An odd number of crossings of the edge states 
at the Fermi level establishes their nontrivial character. (d) Band gaps of slabs with various thicknesses and Bi concentrations $x$. 
Shaded region shows slabs which are non trivial and predicted to provide new candidates for realizing 2D TIs. } 
\end{figure}

The existence of gapless edge states is the hallmark of 2D topological insulators. Thus to verify the topological character of 
2D thin films of Ge(Bi$_x$Sb$_{1-x}$)$_2$Te$_4$, we also computed the edge state energy dispersion for inversion symmetric($x$ = 1.0) and inversion asymmetric($x$ = 0.8) slabs.
As an example, the unit cell for edge state computations of GBT124 ($x$ = 1.0) is shown in Fig. 5(a), where the two edges considered are marked with pink and green vertical lines.
The edge state energy dispersion in GBT124 ($x$ = 1.0) for 14L [Fig. 5(b)] shows that there are three Fermi level crossings, labeled by $\{1,2,3\}$. This odd  
number of crossings between two time-reversal-invariant points confirms the nontrivial nature of these films.
We also computed the edge states of Ge(Bi$_x$Sb$_{1-x}$)$_2$Te$_4$ with $x$ = 0.8 for 21L and the results are shown in Fig. 5(c). Since these slabs are asymmetric under 
inversion, the states associated with opposite edges are not degenerate. The labels $\{1,2,3\}$ and $\{1',2',3'\}$ in Fig. 5(c) indicate the Fermi level crossings from states related to the left and right edges, 
respectively. In this case also the edge states cross the Fermi level three times, implying that 
these thin films are topologically non-trivial.

In order to explore topological phases of 2D Ge(Bi$_x$Sb$_{1-x}$)$_2$Te$_4$ films, we carried out calculations for 7L, 14L, 21L, 28L, and 35L slabs for $x$ = 0.0, 0.2, 0.4, 0.5, 0.6, 
0.8, and 1.0. The topological nature of the films was determined by the methods already described above.
The computed variation of the band gap at the $\overline\Gamma$-point for slabs of different thicknesses and compositions is summarized in Fig. 5(d). The shaded area in the figure identifies slabs with non-trivial 
character. All slabs at x = 0.0 (GST124) are topologically trivial. Excepting the 7L film,
thin films for $x$ = 0.2, 0.4, and 0.5 are metallic, where the metallic character decreases with increasing Bi concentration. An insulating phase is found for slabs with larger values of 
$x$. In particular, we predict that the 28L and 35L films for $x$ = 0.6, 21L, 28L, and 35L films for $x$ = 0.8, and 14L, 21L and 28L films with $x$ = 1.0 are topologically non-trivial. 
Thus, these slabs are possible candidates for realizing 2D TIs, and their solid solutions could realize a 2D TPT.

\section{Summary and Conclusions}
We have investigated electronic structures of thin films of Ge-based system Ge(Bi$_x$Sb$_{1-x}$)$_2$Te$_4$ over the full range of Bi concentrations $x$ within the framework of the density 
functional theory. By using parity analysis and surface state computations, we show that the $x$=1.0 bulk compound, GBT124, is a topological insulator with $Z_2$ = 1;(111) harboring a single 
metallic Dirac cone surface state at the $\overline\Gamma$ point lying within the bulk energy-gap. On the other extreme, at $x$ = 0, the material (GST124) transforms into a standard insulator 
without metallic surface states. Computations as a function of $x$ indicate that a topological phase transition (TPT) takes place for $x$ values between 0.5 and 0.6. Our analysis further 
suggests that a 2D-TPT could be realized in thin films of Ge(Bi$_x$Sb$_{1-x}$)$_2$Te$_4$ by varying Bi concentration. We predict that 28L and 35L films at $x$ = 0.6, 21L, 28L, and 35L films at
$x$ = 0.8, and 14L, 21L and 28L films at $x$ = 1.0, would be 2D TIs. Our study opens up the possibility of identifying the QSH state in thin films of a large materials family, along with that 
of realizing a 2D topological phase transition. 

\section*{ACKNOWLEDGMENTS} 
This work was supported by the Department of Science and Technology, New Delhi (India) through project SR/S2/CMP-0098/2010. The work at Northeastern University was supported by the US 
Department of Energy, Office of Science, Basic Energy Sciences Contract No. DE-FG02-07ER46352, and benefited from Northeastern University's Advanced Scientific Computation Center (ASCC), 
theory support at the Advanced Light Source, Berkeley, and the allocation of time at the NERSC supercomputing center through DOE Grant No. DE-AC02-05CH11231. H.L. acknowledges the Singapore
National Research Foundation (NRF) for support under NRF Award No. NRF-NRFF2013-03.


\begin{thebibliography}{10}
\bibitem{footnoteC}
Note that we have modeled non-stoichiometric compositions by using appropriate supercells of ordered phases. It will be interesting to consider disorder effects using first-principles approaches within the DFT framework (Ref.~\onlinecite{disorder}), although virtual crystal approximation (Ref.~\onlinecite{VCA}) might provide a reasonable scheme for describing effects of Bi/Sb substitution.
\bibitem{footnoteA} Since the DFT correctly predicts that Ge(Bi$_x$Sb$_{1-x}$)$_2$Te$_4$ is a trivial insulator at x=0 and a non-trivial insulator at  x=1, we can have confidence in the prediction that a transition will take place for some value of x over the range 0-1, uncertainty in estimating the precise transition point due to inherent limitations of the DFT in underestimating gap sizes notwithstanding.
\bibitem{footnoteB} Figure 3(b) indicates that the error in estimating the transition point will depend on details of how the corrections to the DFT affect the valence and conduction levels. For example, if the positions of the conduction and valence levels should undergo similar corrections (compared to the DFT) as a function of x, we may see relatively little change in the transition point in Fig. 3(b).
\end{thebibliography}
\end{document}